# Distances and Evolutionary States of Supernova Remnant G18.9-1.1 and Candidate G28.6+0.0

**Sujith Ranasinghe[1*], Denis Leahy[1], Wenwu Tian[1,2]**

[1]Department of Physics & Astronomy, University of Calgary, Calgary, Canada
[2]National Astronomical Observatories, CAS, Beijing, China
Email: *syranasi@ucalgary.ca

## Abstract

HI spectra of the supernova remnant G18.9-1.1 and the supernova remnant candidate G28.6+0.0 are analyzed. We compared the spectra to $^{13}$CO emission spectra and to spectra of HII regions in the area to determine kinematic distances. G18.9-1.1 is at $2.1 \pm 0.4$ kpc and G28.6+0.0 is at $9.0 \pm 0.3$ kpc from the Sun. Using the published X-ray spectra of G18.9-1.1, we apply supernova remnant models for shocked-ISM temperature and emission measure. We find that G18.9-1.1 has low, but not atypical, explosion energy of $\approx 3 \times 10^{50}$ erg and is in a low-density region of the ISM, ~0.1 cm$^{-3}$. It has age ~3700 yr if the ejecta mass is 1.4 $M_\odot$, typical of Type Ia SNe, or ~4700 yr if the ejecta mass is 5 $M_\odot$ typical of core-collapse SN. The candidate G28.6+0.0 does not have reported X-ray emission, so we apply a basic Sedov model. The Sedov age is ~600 yr if the ISM density is 1 cm$^{-3}$ but could be as old as ~6000 yr if the ISM density is as high as 100 cm$^{-3}$.

## 1. Introduction

Supernova remnants (SNRs) remain an important area of study due to their role in interstellar medium (ISM) and Galactic evolution. To determine the basic parameters (e.g. age and size) of a SNR, a distance estimation is necessary. There are ~294 Galactic SNRs [1], and ~62 SNRs in the area covered by the VLA Galactic Plane Survey (VGPS). Distances to 28 of the radio-bright SNRs[1] were determined by previous work (see [2] [3] [4] and [5]). Two more distances to SNRs were estimated by [6], using data from the HI, OH, Recombination line survey

[1]Two are no longer classified as SNRs ([1] [12]).

of the Milky Way (THOR). Furthermore, the evolutionary state of 15 SNRs was determined by [7].

For this work, we chose radio-faint SNRs that were omitted from the previous selection, by carefully investigating their HI channel maps. Due to its diffuse nature, determining the distance to the SNR G18.9-1.1 using HI absorption spectra has been a difficult task. First detected by [8], the morphology, spectrum and presence of the center bar directed it to be classified as a composite SNR ([9] [10]). Reference [11] determined a radio spectral index of $-0.39 \pm 0.03$ ( $S_\nu \propto \nu^\alpha$ ) for the SNR.

The SNR candidate G28.6+0.0 (G28.56+0.00 in [12]) is located in the G28.6-0.1 complex which consists of HII regions and radio sources. Initially [13] suggested a shell-like object to be an SNR due to its non-thermal radio spectrum. However, [14] presented an X-ray analysis of the SNR and found that it emits predominantly synchrotron X-rays from the shell and retained the original SNR ID. Reference [13] argued that the integrated flux density at 20 cm for G28.6+0.0 was clearly higher than the 6 cm and 3 cm flux densities to suggest the non-thermal nature of the SNR candidate. Using THOR data and the 1400 MHz radio continuum data from the VGPS data along with the Spitzer GLIMPSE 8.0 μm and Spitzer MIPSGAL 24 μm data, [12] classified the object as a SNR candidate.

In this paper, we present the HI and $^{13}$CO data and analysis in Section 2. We derive distances for the SNR and SNR candidate in Section 3. Evolutionary models for the two objects are presented in Section 4.

## 2. Data and Analysis

The basic parameters of the SNR G18.9-1.1 and the SNR candidate G28.6-0.0, were obtained from the catalogue of Galactic SNRs [1] and the catalogue of High Energy Observations of Galactic Supernova Remnants (http://www.physics.umanitoba.ca/snr/SNRcat/ [15]).

### Radio and $^{13}$CO Emission Data

The 1420 MHz and HI data were retrieved from the [16] and the $^{13}$CO spectral line from the Galactic Ring Survey of the Five College Radio Astronomical Observatory (FCRAO) 14 m telescope [17]. We followed the method of HI absorption spectra construction, spectral and error analysis as outlined by [18] and [2].

We used MEANLEV, a software program in the DRAO EXPORT package to construct HI absorption spectra. A main advantage of MEANLEV is that, the "on" (source) and "off" (background) spectra could be extracted by user-specified threshold continuum brightness ( $T_B$ ) levels. Once a spatial region, normally a box is selected, the source and background spectra are extracted using pixels above (for source) and below (for background) the given $T_B$ level. This maximizes the contrast difference in ( $T_B$ ), which maximizes the signal-to-noise in the HI spectrum.

The THOR survey covers an area between the Galactic longitudes 14.5° and 67.4˚ and latitudes −1.25˚ and 1.25˚ with a continuum angular resolution of 25″ and HI data angular resolution of 40″ and spectral resolution of 1.6 km·s⁻¹ [19]. We constructed HI emission spectra using the THOR data for the same source (on) and background (off) regions as chosen for the VGPS spectra. This was done in order to assess whether we could obtain an improvement of the spectra. However, we found there to be almost no difference between the THOR and the VGPS spectra: the prominent emission and absorption features of the source regions were present in the spectra from both surveys. Even though the THOR data has a higher angular resolution, the spectral resolution is comparable to the VGPS data. Thus, for this analysis, we found that the VGPS data was sufficient.

To estimate the kinematic distances to SNRs, a reliable rotation curve is necessary. For this work, we adopt the universal rotation curve (URC) of [20] with the [21] parameters, where the Galactocentric radius of $R_0$ = 8.34 ± 0.16 kpc and orbital velocity of the sun $V_0$ = 241 ± 8 km·s⁻¹.

## 3. Results

### 3.1. SNR G18.9-1.1

The diffuse nature of the SNR G18.9-1.1 has made extracting absorption spectra quite difficult, and care needs to be taken to avoid being dominated by noise. An inspection of the HI channel maps by [10] found a depression at 18 km·s⁻¹ giving a distance of 2 kpc to the SNR. However, [10] doesn't present HI absorption spectra. Using red clump stars [22] estimated the distance to be 1.8 ± 0.2 kpc, an estimation consistent with [10].

The SNR consists of two brighter arcs (A & B in **Figure 1**). We extracted multiple spectra for both regions and found that arc B's brightest region produces the best absorption spectrum. **Figure 2** shows the HI emission spectra of the source region and the background region (top panel). The bottom panel shows the HI absorption spectrum. The HI channel maps reveal that the absorption/emission features seen in the negative velocity range are likely dominated by noise. There is no clear evidence of absorption up to the tangent point.

The absorption features seen at ~40 and ~95 km·s⁻¹ are false features that have HI emission features coincidentally in the background region. The only real absorption feature is seen at ~18 - 23 km·s⁻¹. This feature appears consistently in both arcs A & B. **Figure 3** is the image of HI channel maps averaged between 18.32 and 23.27 km·s⁻¹. This shows the absorption matching the continuum morphology of arcs A and B, confirming that the absorption is real. We have used different combinations of the HI channel maps to determine the velocity range where the absorption occurs. Using the velocity 23 km·s⁻¹ of the upper edge of the absorption feature, we find the SNR is at a distance of 2.1 ± 0.4 kpc.

### 3.2. SNR Candidate G28.6+0.0

We present the 1420 MHz radio continuum image of SNR candidate G28.6+0.0 including the HII regions and radio sources of the G28.6-0.1 complex in **Figure 4**.

The labelling convention is the same as that in [13]. The HII region labelled as D by [13] is in fact two separate HII regions G028.610+0.020 and G028.600-0.011 with the radio recombination line (RRL) velocities of 96.6 and 41.9/92.8 km·s⁻¹, respectively [23]. For this work, we re-label the two HII regions as D and D' (see

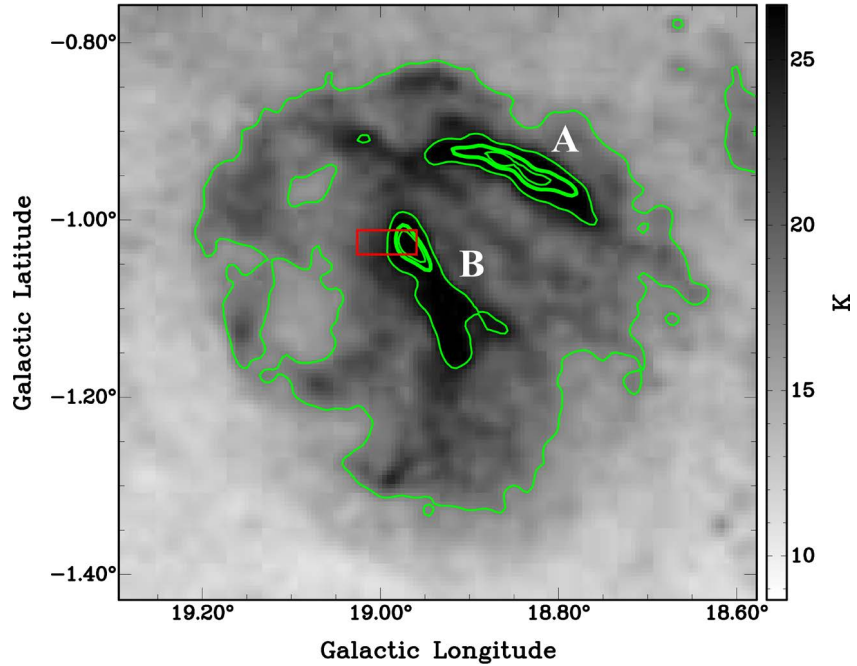

**Figure 1.** VGPS 1420 MHz Continuum image of G18.9-1.1 with green contours at continuum brightness temperatures of 17.5, 25, 30 and 32 K. The red box denotes the region used for the HI absorption spectrum extraction.

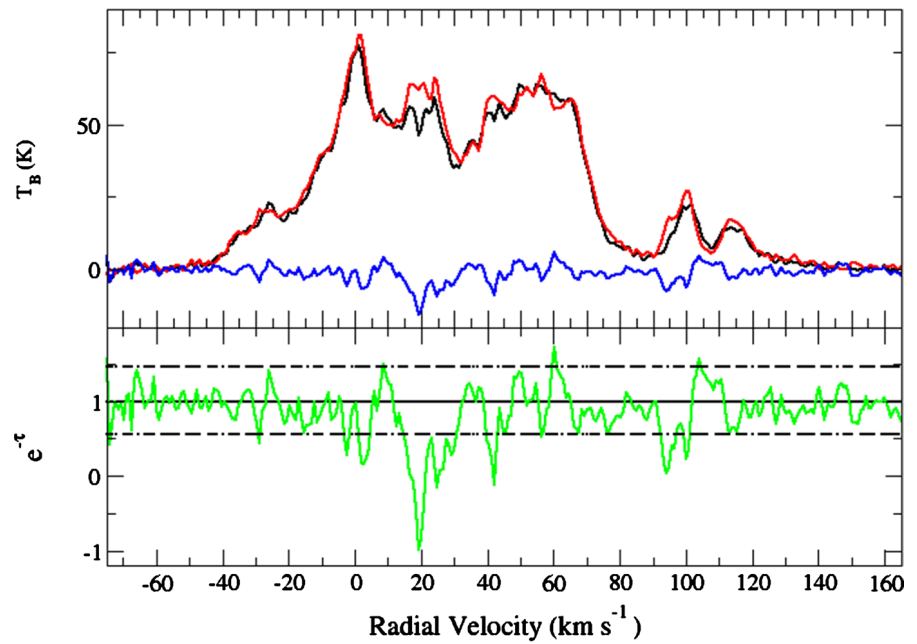

**Figure 2.** G18.9-1.1 absorption spectrum. Top panel: emission spectra for the source (black), background (red) and difference (blue). Bottom panel: absorption spectrum (green) and the $2\sigma$ noise level (Dashed) of the HI absorption spectrum.

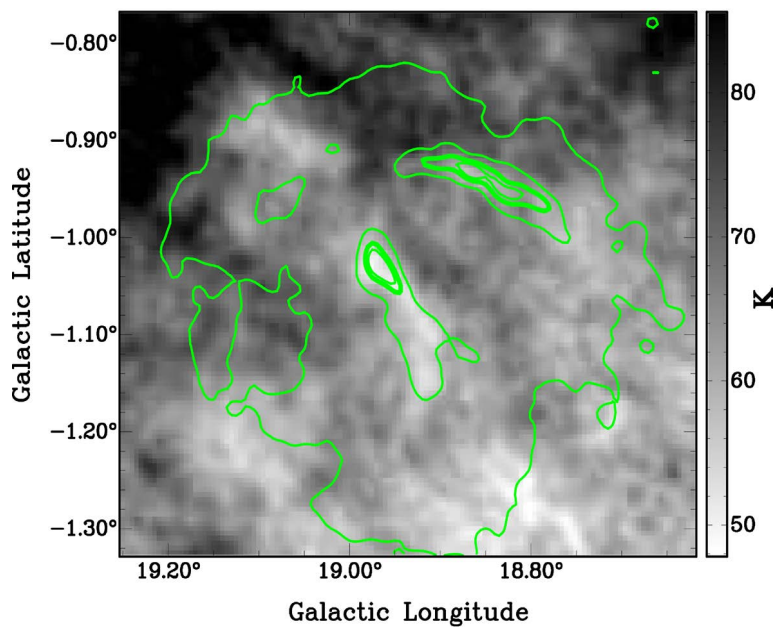

**Figure 3.** Combined HI channel maps +18.32 to +23.27 km·s⁻¹ for the SNR G18.9-1.1. 1420 MHz continuum contour lines are in green at 17.5, 25, 30 & 32 K.

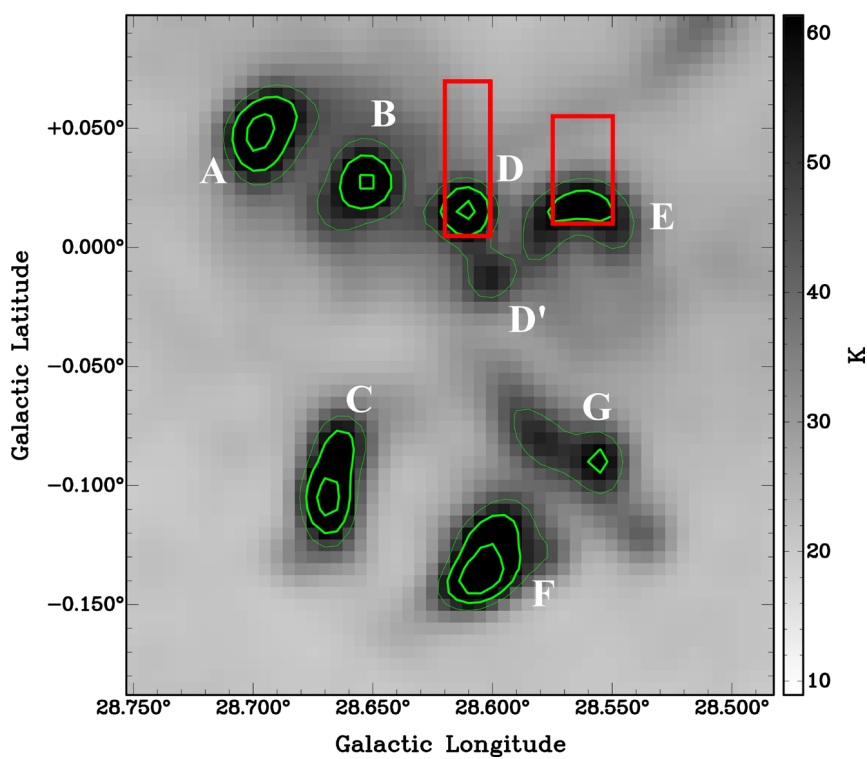

**Figure 4.** VGPS 1420 MHz Continuum image of the G28.6-0.1 complex, with green contours at continuum brightness temperatures of 48, 60 and 80 K. Red boxes denote the regions used for the HI absorption spectrum extraction.

**Figure 4**). We extracted HI absorption spectra for the SNR candidate and HII region (D), which are shown in **Figure 5**. The scaled ¹³CO emission spectra of the background (black curve) and source (purple curve) regions are shown in the bottom panels of Figure 5.

The HI absorption feature seen in the negative velocity range (near −40 km·s⁻¹) is false, because HI emission is found coincidentally in the chosen background region (Figure 6). Furthermore, this feature is seen in both the SNR candidate and HII region spectra. For both sources, HI absorption is seen up to the tangent point, which yields a lower limit distance of 7.3 kpc (the tangent point distance). ¹³CO emission features are seen at the velocities ~40, ~75 and

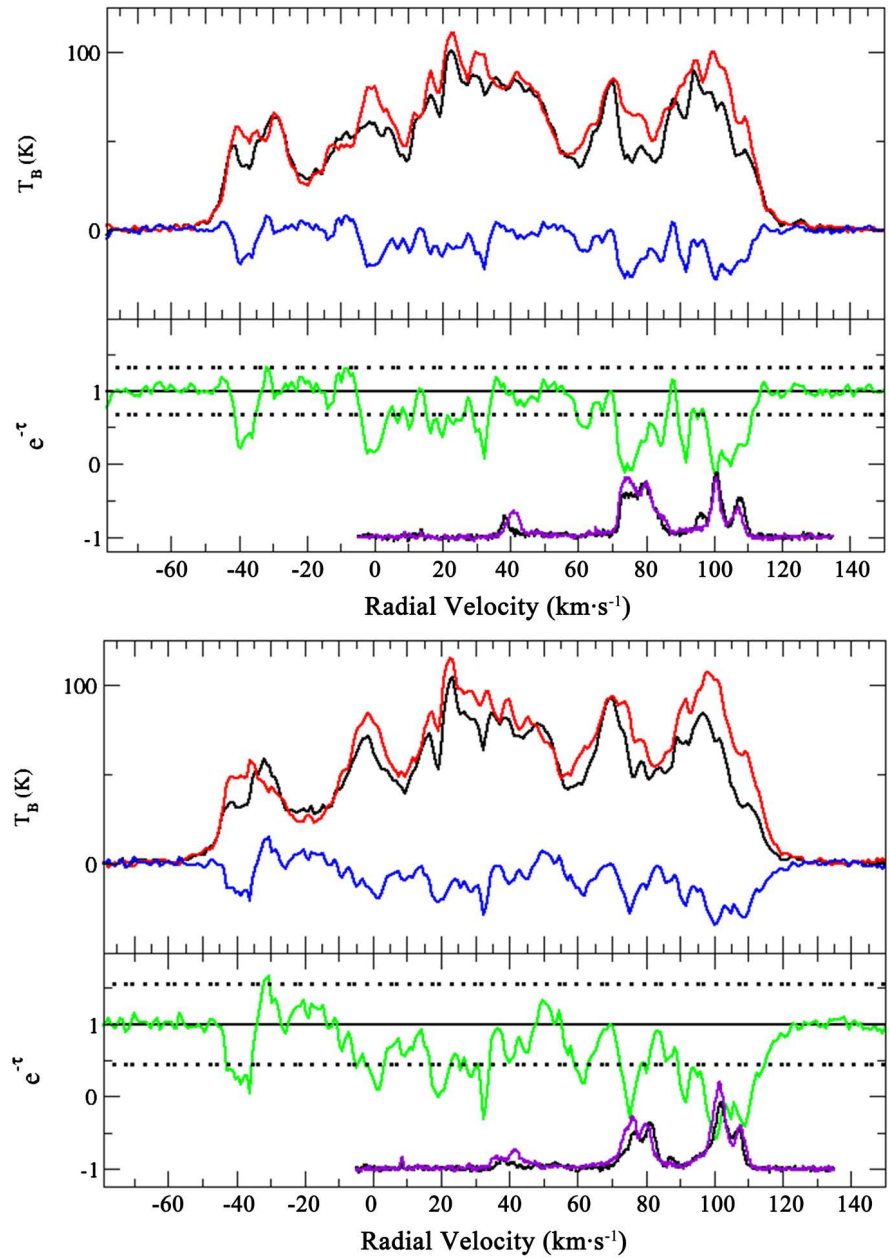

**Figure 5.** HI absorption spectra for the SNR candidate G28.6+0.0 (top) and the HII region G028.610+0.020 (bottom). For both panels, HI emission spectra for the source are in black, HI background spectra are red and the difference spectra are blue. The HI absorption spectra are green, their 2σ noise levels are the dashed lines. The scaled ¹³CO emission spectra of the background are in black and of the source are in purple.

~100 km·s⁻¹, and are consistent with the HI absorption features for both objects. Thus, the ¹³CO clouds at these velocities are in front of the objects.

The absorption spectra of both the SNR candidate and HII region G028.610 +0.020 are nearly identical: the differences are less than the noise level (see **Figure 5**). This has been verified by examining HI channel maps. The HII region G028.610+0.020 has a RRL velocity of 96.6 km·s⁻¹ and is at a distance of 9 kpc (D in **Table 1**). It is likely that the SNR candidate is related to the HII region because the spectra are so similar. We deduce that the SNR candidate is at the same distance of 9.0 ± 0.3 kpc.

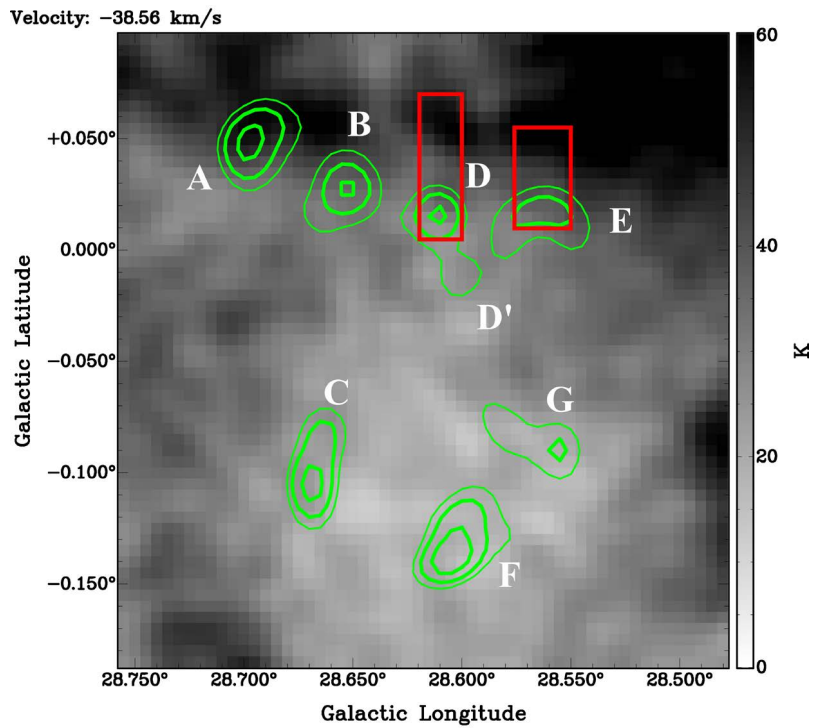

**Figure 6.** VGPS HI channel map of G28.6-0.1 complex at −38.56 km·s⁻¹. The dark shading in the upper parts of the red boxes indicates coincidental HI emission in the background region, which yields a false absorption feature. The green contours are at continuum brightness temperatures of 48, 60 and 80 K.

**Table 1.** Distances and radial velocities of the objects in G28.6-0.1 complex.

| Source | VLSR[a] (km·s⁻¹) | R[b] (kpc) | d[b] (kpc) | Refs[b] | R[c] (kpc) | d[c] (kpc) |
|--------|------------------|------------|------------|---------|------------|------------|
| A | 100.7[R] | 4.3 | 7.5 | [25] | 4.2 | 8.6 |
| B | 102.4[R] | 4.3 | 8.7 | [24] | 4.2 | 8.5 |
| C | 86[H] | 4.6 | 9.6 | [4] | ... | ... |
| D | 96.6[R] | 4.4 | 5.8 | [23] | 4.3 | 9.0 |
| D' | 41.9/92.8[R] | 6.0 | ... | [23] | 6.1/4.4 | 11.9/9.2 |
| E | ... | ... | ... | ... | 4.3 | 9.0 |
| F | 86[H] | 4.6 | 9.6 | [4] | ... | ... |
| G | 86[H] | 4.6 | 9.6 | [4] | ... | ... |

[a]The superscripts R and H indicates RRL and HI absorption velocities respectively. [b]Literature values of Galactocentric radius (R) and distance (d). [c]Re-calculated Galactocentric radius (R) and distance (d) using the [21] rotation curve.

## 4. Discussion

We derive a distance to the SNR G18.9-1.1 of 2.1 ± 0.4 kpc, because of the observed absorption up to but not beyond the velocity of 23 km·s$^{-1}$. In some cases, lack of absorption beyond a particular velocity could be interpreted as lack of absorbing HI at higher velocities. However, in this case, the HI emission at higher velocities is comparable to the emission at ~23 km·s$^{-1}$. Thus, if the SNR was located beyond ~23 km·s$^{-1}$, it would show HI absorption.

Furthermore, we examined the absorption spectra of HII regions 19.050 - 0.593 and 18.881 - 0.493 which are ~30′ and ~35′ from the SNR center, respectively. The continuum brightness temperature of the HII region 19.050 - 0.593 has maximum of 47.9 K which is comparable to the maximum brightness temperature of the SNR at 34.9 K. The RRL velocity of the HII region is 68.2 ± 1.3 km·s$^{-1}$ [24]. From the HI channel maps, it is seen that strong absorption features are present at 53.78 km·s$^{-1}$ and weaker absorption features at 68.62 km·s$^{-1}$, consistent with the RRL velocity. Using the Galactic rotation curve, the HII region is at a distance of 4.3 ± 0.2 kpc. The HII region 18.881 - 0.493 has a RRL velocity of 65.5 ± 0.8 km·s$^{-1}$. Strong absorption for 18.881 - 0.493 is seen up to 70.26 km·s$^{-1}$ and the HI absorption velocity is consistent with the RRL velocity within the radial velocity error of 5.3 km·s$^{-1}$ [4]. 18.881 - 0.493 is at a distance of 4.2 ± 0.2 kpc. Thus, there is good evidence for absorbing HI at velocities higher than the observed absorption in SNR G18.9-1.1. We conclude the SNR is at the kinematic distance of 2.1 - 0.4 kpc, and consistent with [10] and [22].

G18.9-1.1 was observed in X-rays with the ROSAT PSPC [26] and ASCA [27]. We adopt a plasma electron temperature of 1.12 keV from [27] and use the average of the GIS lower emission measure (EM) norm limit and the ROSAT upper EM norm limit together with our distance to estimate a value for EM of 2.6 × 10$^{57}$ cm$^{-3}$. With the measured radius, electron temperature and emission measure, we apply the SNR models given in [28].

These models yield the age, explosion energy and ISM density that result in a SNR with the observed parameters. **Table 2** lists the results of our modelling of G18.9-1.1. The SNR age is intermediate, between 2500 and 5000 yr, so that electron heating by ion collisions not yet complete. Assuming $T_e = T_i$ (*i.e.* $T_e$ too large) gives a too-large age and too-low explosion energy, in effect making the SNR older and weaker than its actual age in an attempt to match the observed electron temperature. We see that the realistic models, with non-zero ejecta mass, yield ages which depend on assumed ejecta mass. The energy and ISM density also depends on ejecta mass but not as strongly. The SNR models also predict the temperature and emission measure of the reverse-shocked ejecta. When better X-ray observations are available, observation of the reverse-shocked ejecta can distinguish between models with different ejecta mass and ejecta density profiles (the value of n).

**Table 2.** SNR models for G18.9-1.1.

| Model | $M_{ejecta}$[a] ($M_\odot$) | Age (yr) | $E_{51}$ ($10^{51}$ erg) | $n_0$ ($cm^{-3}$) |
|---|---|---|---|---|
| Sedov/ $T_e = T_i$[b] | 0 | 4000 | 0.13 | $6.8 \times 10^{-2}$ |
| Sedov | 0 | 3300 | 0.27 | $9.5 \times 10^{-2}$ |
| WL[c] | 0 | 2600 | 0.45 | $2.6 \times 10^{-2}$ |
| Standard/ $s = 0$, $n = 7$[d] | 1.4 | 3700 | 0.28 | $9.2 \times 10^{-2}$ |
| Standard/ $s = 0$, $n = 7$[d] | 5.0 | 4700 | 0.26 | $8.0 \times 10^{-2}$ |

[a]Ejecta mass, $M_{ejecta}$, is an input value. [b]Assumes $T_e = T_i$. All other models include electron heating by Coulomb collisions. [c]WL: cloudy ISM model of [29] with C/$\tau$ = 4. [d]Standard model of [28], based on unified SNR evolution of [30]. s = 0 is for uniform ISM, n = 7 is for ejecta density profile $\propto r^7$.

For SNR candidate G28.6+0.0, we measure its diameter as 2.9 arcmin using the VGPS radio continuum image. Our derived distance then gives its radius as R = 3.8 pc. Because there is no reported X-ray emission, we cannot apply models which use the X-ray emission to determine explosion energy $E_0$, ISM density $n$, and age [28]. Thus, we apply a basic Sedov model with the assumption of $E_0 = 0.5 \times 10^{51}$ erg. This is the mean value found for Galactic SNRs and for LMC SNRs ([7] [31]). Then we take trial values of ISM density $n = 0.1, 1, 10$ and $100$ $cm^{-3}$ to obtain ages of 180, 580, 1800 and 5800 yr, respectively. In short, the small observed radius implies a small age unless the density is high.

## 5. Summary

In this work, we analyze HI spectra of SNR G18.9-1.1 and SNR candidate G28.6+0.0. $^{13}CO$ emission spectra and HI spectra of HII regions in the area are compared to the spectra of our objects to verify their kinematic distances. The distance of G18.9-1.1 is 2.1 ± 0.4 kpc and of G28.6+0.0 is 9.0 ± 0.3 kpc.

Our derived radius of G18.9-1.1, with published X-ray shocked-ISM temperature and emission measures, allow application of SNR models. The models show that G18.9-1.1 has slightly low explosion energy of ~3 × $10^{50}$ erg and is in a low-density region of the ISM, ~0.1 $cm^{-3}$. If the ejecta mass is 1.4 $M_\odot$, typical of Type Ia SNe, its age is ~3700. If the ejecta mass is 5 $M_\odot$, typical of core-collapse SN, its age is ~4700 yr. We apply a basic Sedov model to SNR candidate G18.9-1.1. If the ISM density is 1 $cm^{-3}$ its Sedov age is ~600 yr, but if the ISM density is as high as 100 $cm^{-3}$ its age is significantly larger, ~6000 yr.

## Acknowledgements

This work was supported by a grant to DL from the Natural Sciences and Engineering Research Council of Canada.

## Conflicts of Interest

The authors declare no conflicts of interest regarding the publication of this paper.